\documentclass[twocolumn]{aastex62}

\graphicspath{{./}{figures/}}
\usepackage{verbatim}
\usepackage[utf8]{inputenc}
\usepackage{hyperref}
\usepackage{units}
\usepackage{colortbl}
\usepackage{newtxtext,newtxmath}
\usepackage{graphicx}	
\usepackage{amsmath}	
\usepackage{amssymb}	

\received{November 15, 2019}
\revised{January 31, 2020}
\accepted{March 06, 2020}
\submitjournal{ApJ}

\shorttitle{Silicate grain growth due to ion trapping in SNRs}
\shortauthors{Kirchschlager, Barlow, Schmidt}

\begin{document}

\title{Silicate grain growth due to ion trapping in oxygen-rich supernova remnants like Cassiopeia~A}

\correspondingauthor{Florian Kirchschlager}
\email{f.kirchschlager@ucl.ac.uk}

\author[0000-0002-3036-0184]{Florian Kirchschlager}
\affil{Department of Physics and Astronomy,\\
University College London,\\
Gower Street,\\
London WC1E 6BT,\\
United Kingdom}

\author{M.~J.~Barlow}
\affiliation{Department of Physics and Astronomy,\\
University College London,\\
Gower Street,\\
London WC1E 6BT,\\
United Kingdom}

\author{Franziska~D.~Schmidt}
\affiliation{Department of Physics and Astronomy,\\
University College London,\\
Gower Street,\\
London WC1E 6BT,\\
United Kingdom}

 
\begin{abstract}
Core-collapse supernovae can condense large masses of dust post-explosion. 
However, sputtering and grain-grain collisions during 
the subsequent passage of the dust through the reverse shock can 
potentially destroy a significant fraction of the newly formed dust before 
it can reach the interstellar medium. Here we show that in oxygen-rich 
supernova remnants like Cassiopeia~A the penetration and trapping within silicate 
grains of the same impinging ions of oxygen, silicon and magnesium that 
are responsible for grain surface sputtering can significantly reduce the 
net loss of grain material. We model conditions representative of dusty clumps (density contrast $\chi=100$)
passing through the reverse shock in the oxygen-rich Cassiopeia~A remnant and find that, 
compared to cases where the effect is neglected, as well as facilitating the formation 
of grains larger than those that had originally condensed, ion trapping 
increases the surviving masses of silicate dust by factors of up to two to four, 
depending on initial grain radii. For higher density contrasts ($\chi\gtrsim180$), we find that the effect of gas accretion on the surface of dust grains surpasses ion trapping, and the survival rate increases to $\unit[{\sim}55]{\%}$ of the initial dust mass for $\chi=256$.
\end{abstract}


\section{Introduction}
\label{101}
The formation of dust in core-collapse supernova ejecta has been established for at least three decades (\citealt{Lucy1989}). Along with SN 1987A and the Crab nebula, Cassiopeia~A (Cas~A) is the most studied dusty supernova remnant (SNR). The remnant of an explosion from around the year 1680, at a distance of ${\sim}\unit[3.4]{kpc}$, has a diameter of ${\sim}\unit[3.4]{pc}$ today (\citealt{Reed1995}). Within Cas~A, dense gas clumps and knots are observed which are associated with the location of freshly produced dust (\citealt{Lagage1996,  Rho2012}). Strong emission features in the mid-IR spectra of Cas~A have been identified with silicaceous grains by \cite{Rho2008}, consistent with the very oxygen-rich composition of its ejecta (\citealt{Chevalier1979}).

The total dust mass in the ejecta has been derived by different observational strategies to be between $\unit[{\sim}0.1]{M_\odot}$ and ${\sim}\unit[1]{M_\odot}$ (\citealt{Dunne2009, Sibthorpe2010,Barlow2010,DeLooze2017,Bevan2017,Priestley2019a}), while a theoretical study of dust formation and evolution predicted Cas~A's dust mass to be of the order of $\unit[0.08]{M_\odot}$ (\citealt{Nozawa2010}). This apparent discrepancy between theory and observations is accompanied by a difference in estimated grain sizes: supernova dust grain radii derived from IR emission fitting as well as from fits to asymmetric optical line profiles are of the order of $\unit[0.1]{\mu m}$ up to a few microns (\citealt{Stritzinger2012,Owen2015,Fox2015,Wesson2015,Bevan2016,Priestley2020}), while dust formation theories predict grain sizes $\lesssim\unit[0.1]{\mu m}$ (\citealt{Todini2001, Nozawa2003, Bianchi2007, Marassi2015, Biscaro2016}).

Aggravating this situation, the interaction of the blast wave with the circumstellar and interstellar medium causes a reverse shock (\citealt{Gotthelf2001}) which can have a crucial effect on the evolution of the embedded dust material.
Significant amounts of dust can be destroyed or transformed when the reverse shock encounters the dusty ejecta clumps (e.g.~\citealt{Nozawa2010, Silvia2010, Bocchio2016, Micelotta2016}). Recently, \cite{Kirchschlager2019b} (hereafter K19) have studied dust survival rates in Cas~A as a function of clump densities, dust material and initial grain size distributions. They found that grain-grain collisions and sputtering are synergistic and that grain-grain collisions have a strong influence on the dust survival rate, by fragmenting larger grains to smaller particles that are more easily destroyed by sputtering.

Ion implementation and trapping in dust grains has not been considered so far as a process which in suitable environments can counteract grain destruction by sputtering. The gas in the ejecta clumps of Cas~A is composed of ${\sim}\unit[80]{\%}$ oxygen with contributions from Ne, Mg, Si, S, Ar, and a negligible amount of hydrogen and helium (\citealt{Willingale2002, Docenko2010}). \cite{Chevalier1979} detected dense knots which are made of almost pure oxygen. Due to the high gas temperatures and shock velocities, energetic oxygen and other heavy ions can penetrate deep into the grains. For grain temperatures below $\unit[{\sim}500]{K}$ the diffusion rate of oxygen atoms in silicates is very low (\citealt{Brady1995}) and thus they will be trapped once they have intruded into the grain. This will automatically lead to grain growth and an increase in dust mass. Therefore, grain growth by ion trapping can potentially counteract destructive processes such as thermal and kinematic sputtering as well as fragmentation and vaporisation in grain-grain collisions. 

In this paper, the  effect of ion trapping on the dust evolution in Cas~A is studied. In Section~\ref{102} we describe our model setup and give a summary of the considered dust processes. We introduce oxygen ion trapping in dust grains in Section~\ref{103}. The results of our study are presented in Section~\ref{sec3} and a summarising conclusion is given in Section~\ref{sec4}.

  \section{Model and methods}
\label{102}
In order to investigate the effect of oxygen ion trapping in dust grains in the ejecta of Cas~A, we performed hydrodynamical simulations using the grid-based code \textsc{AstroBEAR} (\citealt{Carroll-Nellenback2013}). \textsc{AstroBEAR} simulations model only the gas phase (Section~\ref{2020a}) and we use our external post-processing code \mbox{\textsc{Paperboats}} (K19) to investigate dust motions as well as dust grain growth or dust destruction in a high-velocity, hot gas environment (Sections~\ref{d3}-\ref{103}). 

\subsection{Cloud-crushing problem}
\label{2020a}
The applied model is based on the setup used in K19 and is briefly summarised here.

Instead of simulating the entire SN ejecta, we consider only a section of it. We focus on the cloud-crushing scenario (\citealt{Woodward1976, Silvia2010}) in which a planar shock is driven into an over-dense spherical clump of gas which is embedded in a low-density ambient medium (see Fig.$\,$2 in K19). The interaction between this reverse shock and the clump is assumed to be of a similar nature for all ejecta clumps (\mbox{${\sim}10^5-10^6$} in total, see K19) so that our results can be applied and projected to them. Following this approach, we are able to investigate the destruction of a single clump at higher spatial resolution. The initial (pre-shock) conditions comprise the clump radius $R_\textrm{cl}=\unit[10^{16}]{cm}$, uniform gas number densities in the ambient medium $n_\textrm{am}=\unit[1]{cm^{-3}}$, while in the clump $n_\textrm{cl}=\chi\,n_\textrm{am}$, where $\chi$ is the density ratio between clump gas and ambient medium gas, along with the gas temperatures in the ambient medium and in the clump, $T_\textrm{am}=\unit[10^4]{K}$ and $T_\textrm{cl}=\unit[10^2]{K}$, respectively. Unless stated otherwise we set $\chi =100$. The shock velocity in the ambient medium is adopted to be $v_\textrm{sh} = \unit[1600]{km/s}$ (\citealt{Micelotta2016}) which transforms to $\unit[{\sim}160]{km/s}$ in the clump, and the mean molecular weight of the pre-shock gas is \mbox{$\mu_\textrm{gas}=16$} which corresponds to pure oxygen gas.

The simulation is executed for three cloud-crushing times  $\tau_\textrm{cc} = \chi^{0.5} R_\textrm{cl}/v_\textrm{sh}$
(\citealt{Klein1994}) after the first contact of the shock with the clump, which gives the characteristic time for the clump to be crushed and evolved by the shock. For the chosen initial conditions, the simulation time is then ${\sim} \unit[61.5]{yr}$. In contrast to K19, we perform not only 2D but also 3D simulations. The computational domain consists of $420\times140$ and $420\times140\times 140$ cells, respectively, representing a box of length \mbox{$l_\textrm{box}=21\,R_\textrm{cl} = \unit[0.068]{pc}$} and width \mbox{$w_\textrm{box}= 7\,R_\textrm{cl} = \unit[0.023]{pc}$}, and  for the 3D simulations with a box height \mbox{$h_\textrm{box} = w_\textrm{box}$}. The post-shock gas conditions are then calculated by \mbox{\textsc{AstroBEAR}} using the Rankine-Hugoniot jump conditions and taking into account radiative cooling for a gas of pure oxygen (see Fig.~3 in K19). 
The shock evolves in the direction of $l_\textrm{box}$ and the given box size ensures that the clump material stays in the domain throughout the entire simulation. 

At this resolution, the 2D and 3D simulations of the cloud-crushing problem show only moderate differences which is why we present here only the 3D results. We expect more subtle variations between 2D and 3D at higher resolution but we are limited to 20 cells per clump radius due to the large computational effort for highly resolved 3D post-processing simulations (Section~\ref{sec_chap23}).

\subsection{Dust model}
\label{d3}
The grains in our study are spherical with grain radius $a$. At the start the grain sizes follow a log-normal distribution (e.g.~\citealt{Nozawa2003}) that is described by the radius $a_\textrm{peak}$ at which the distribution has its maximum and the dimensionless quantity $\sigma$ that defines the width of the distribution. We fix $\sigma=0.1$ and vary  $a_\textrm{peak}$ to $0.01, 0.1$, and $\unit[1]{\mu m}$. The grain size distribution spans from $a_\textrm{min,abs}=\unit[0.6]{nm}$ to $a_\textrm{max,abs}=\unit[10]{\mu m}$ and is discretised in 40 log-spaced size bins. The dust material is silicate and the material parameters required for the dust post-processing are given in Table~2 of K19. Initially, the dust is at rest compared to the clump gas and  homogeneously distributed in the clump with a gas-to-dust mass ratio of \mbox{$\Delta_\textrm{gd} = 10$} (\citealt{Priestley2019a}) while the ambient medium is dust-free.

\subsection{Dust processing}
\label{sec_chap23}
We use our parallelised 3D external dust-processing code \mbox{\textsc{Paperboats}} (K19) to determine the dust evolution in the shocked clump based on the time- and spatially-resolved gas density, velocity and temperature output of \textsc{AstroBEAR} as well as on the initial dust conditions (Section~\ref{d3}). We give here a short overview of the processes considered and refer to K19 for a detailed description. 

\mbox{\textsc{Paperboats}} was developed to simulate dust motions in the shocked clump as well as dust destruction and grain growth processes. The dust is accelerated by the streaming gas taking into account both collisional and plasma drag (\citealt{Baines1965, Draine1979}). Destruction processes include fragmentation and vaporisation in grain-grain collisions (e.g. \citealt{Tielens1994, Jones1996}) as well as thermal and non-thermal (kinematic) sputtering (e.g. \citealt{Barlow1978,Bocchio2014}). As the dust temperatures in Cas~A are of the order of or even below $\unit[100]{K}$ (\citealt{DeLooze2017, Priestley2019a}), dust destruction by evaporation can be neglected.

Besides the dust destruction processes, two grain growth processes were also previously considered: the coagulation of dust grains and the accretion of (dusty) gas onto the surfaces of the grains. Both can occur, either at low relative velocities in a grain-grain collision or when the energy $E$ of an impinging gas particle is below the threshold energy $E_\textrm{sp}$ for the sputtering of grain atoms (\citealt{Bohdansky1980, Tielens1994}). For the latter case, the gas particle can be accreted in a process akin to negative sputtering. Sticking in grain-grain collisions has a negligible effect on the dust processing, as the present grain velocities are mostly too high (K19). 

On the other hand, when the ion impact energy is sufficiently high, gas particles penetrate into the dust grains and can be trapped. This potential grain growth scenario has not been considered so far for the dust evolution in oxygen-rich SNRs and will be outlined in the following section.

\subsection{Oxygen ion trapping}
\label{103}
We have added oxygen ion trapping in dust grains as an additional feature to the sputtering routine of \mbox{\textsc{Paperboats}}. The sputtering models commonly used in the astronomical literature (e.g. \citealt{Tielens1994,Nozawa2006}) only describe the number of detached dust atoms per incident gas particle but do not consider the further progress of the incident gas particle.

Due to the energetic ion bombardment that silicate grains experience in the reverse shock, they should be rendered partially disordered, enabling impacting and subsequently trapped gas atoms of O, Si, Mg, etc. to form new silicate structure bonds with the surrounding grain material atoms. Along with 10-$\mu$m silicate emission features, the oxygen-rich SNRs Cas~A and 
G54.1-0.3 both exhibit an unusual 21-$\mu$m dust emission feature that has been attributed to a number of silicaceous grain materials, including SiO$_2$ and MgSiO$_3$ (\citealt{Rho2008, Rho2018}) and Mg$_{0.7}$SiO$_{2.7}$ (\citealt{Temim2017}). The constituent atoms of each of those materials have a mean atomic weight of $\mu_\textrm{dust}=20$ (in atomic mass units $m_\textrm{amu}$). From its X-ray spectrum \cite{Willingale2002} derived gas-phase abundance number ratios for Cas~A of Si/O = 0.34 and Mg/O = 0.063, corresponding to a mean atomic weight of $19.3$. Heavier species, such as Ca and Fe, will raise the mean atomic weight of impacting ions further. So for such oxygen-rich SNRs we can treat the mean atomic weight of implanted atoms and sputtered atoms as being the same. For convenience, hereafter the trapping  of ions of various elements will be referred to as oxygen trapping, the most abundant element.
  
\subsubsection{Penetration depth}
\label{1033}  
When the energy of the gas particle is high enough it can penetrate into the dust grain. The penetration depth $r_\textrm{p}$ of the gas particle into the dust grain can be calculated using the Bethe-Bloch formalism (see Section 4.6.6 in K19) and depends, beside the particle energy, on the dust material parameters and the ionisation degree of gas and dust grain. Oxygen is the predominant gas species in Cas A's ejecta and has a diffusion timescale (per nanometre) that is much longer than the Hubble time in olivine or quartz grains with dust temperatures of the order of ${\sim}\unit[500]{K}$ or less (\citealt{Brady1995}).

We set the minimum penetration depth that is required to enable oxygen trapping to $r_\textrm{min}=5$ layers of silicate particles  ($\unit[{\sim}1.1]{nm}$). This number is an assumption, however we have tested lower limits of $r_\textrm{min}=0,3,10$ and 20 silicate layers and found that the differences in the net yield are negligible for $r_\textrm{min}\le\unit[5]{layers}$ while $r_\textrm{min}=10$ and $\unit[20]{layers}$ show a significantly reduced oxygen trapping. We note that a minimum penetration depth of 5 silicate layers corresponds to an energy of $\unit[{\sim}30]{eV}$ (which is similar to the sputtering threshold energy $E_\text{sp}$) for a fully ionised oxygen ion. An impinging oxygen ion with $\unit[160]{km/s}$ corresponds to $\unit[2.1]{keV}$ and a penetration depth of $\unit[{\sim}70]{nm}$ (see Fig.~10 in K19 for the relation between penetration depth and ion energy).

On the other hand, the penetration depth of the oxygen particle has to be below the extent of the grain. The geometrically-averaged path length through the grain is given by $ r_\textrm{max}=4/3$ grain radii, which we use as an upper limit.

\subsubsection{Net yield}
\label{104}
The yield $Y$ is the change of the number of dust grain atoms per normal incident gas particle. A positive yield represents dust destruction by sputtering while a negative yield implies dust grain growth by gas accretion (low energy) or ion trapping (high energy).

\begin{figure}
   \centering
    \includegraphics[trim=1.1cm 1.95cm 0.5cm 1.75cm, clip=true,width=1.0\linewidth, page=1]{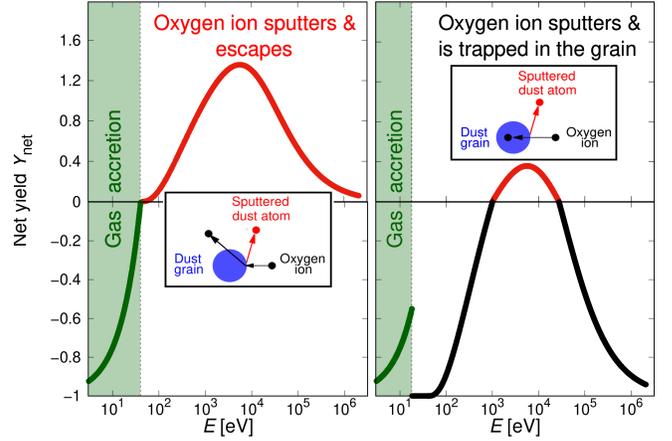} 
     \caption{Net yield $Y_\textrm{net}$ which gives the change of number of atoms in the grain per incident oxygen ion as a function of energy $E$ of the incident oxygen ion. A positive yield corresponds to dust destruction, a negative yield to dust growth. When the energy of the incident oxygen ion is higher than  the sputtering threshold energy, $E_\textrm{sp}$, dust atoms can be sputtered. \textit{Left:} Trapping conditions are not fulfilled (oxygen particle escaping after the sputtering event) so the resulting net yield is equal to the regular sputtering yield, $Y_\textrm{net}=2\,Y_\textrm{sp}$ (red solid line). \textit{Right:} The oxygen ion is trapped so the net yield amounts to $Y_\textrm{net}=2\,Y_\textrm{sp}-1$ (black solid line) which can become negative as less than 1 silicate atom is detached per sputtering event. For both cases, the incident oxygen atom is accreted at lower energies (green shaded region).}
     \label{res_dust_evo} 
  \end{figure}

When the conditions for oxygen ion trapping are fulfilled ($r_\textrm{min}\le r_\textrm{p}\le r_\textrm{max}$), the oxygen ions cause a negative yield \mbox{$Y_\textrm{trap}=-1$.} The net yield resulting from sputtering ($Y_\textrm{sp}$) and trapping ($Y_\textrm{trap}$) amounts to $Y_\textrm{net}=2\,Y_\textrm{sp}-1$ where a factor of two is considered to allow for higher measured sputtering yields at non-normal incidence  over normal-incidence sputtering yields (e.g.~\citealt{Tielens1994}). $Y_\textrm{sp}$ is usually below 1 for oxygen particles impacting on various dust materials (silicate, carbon etc., see \citealt{Tielens1994}), so the net yield can become negative (\textit{right} panel of Fig.~\ref{res_dust_evo}).  If the oxygen particle is not trapped by the grain the net yield amounts to the regular sputtering yield, $Y_\textrm{net}=2\,Y_\textrm{sp}$ (\textit{left} panel of Fig.~\ref{res_dust_evo}).

For energies lower than the sputtering energy threshold $E_\textrm{sp}$ and lower than the energy at which trapping can occur, gas particles are assumed to be accreted onto the grain
surface with a yield \mbox{$Y_\textrm{acc} = -(1-E/E_\textrm{sp})$} (K19). The linear decay of the gas accretion yield enables a continuous transition from the gas accretion to the sputtering regime at the threshold energy $E_\textrm{sp}$ (\textrm{left} panel of Fig.~\ref{res_dust_evo}). When the energy of the impinging oxygen particle is below $E_\textrm{sp}$ but still high enough that it can be trapped, the net yield is set to $Y_\textrm{net}=-1$.

Finally, the resulting net yield amounts  to
\begin{align}
Y_\textrm{net} = \begin{cases}
           2\,Y_\textrm{sp},    &\textrm{if}\hspace{0.1cm}  (E>E_\textrm{sp})\wedge \textrm{not trapped},\\
           2\,Y_\textrm{sp}-1,  &\textrm{if}\hspace{0.1cm}(E>E_\textrm{sp})\wedge \textrm{trapped},\\
	   -1,             &\textrm{if}\hspace{0.1cm}	      (E<E_\textrm{sp})\wedge \textrm{trapped},\\           
	   Y_\textrm{acc},   &\textrm{if}\hspace{0.1cm} (E<E_\textrm{sp})\wedge \textrm{not trapped}.\\	   
          \end{cases}
\end{align}

\subsubsection{Gas depletion factor}
Sputtering, gas accretion and gas trapping cause a change of the grain size $a$ to $a-\left(\frac{\textrm{d}a}{\textrm{d}t}\right)\Delta t$ during the time-step $\Delta t$, where 
\begin{align}
 \frac{\textrm{d}a}{\textrm{d}t} = \frac{\mu_\textrm{dust} m_\textrm{amu}}{4\,\rho_\textrm{bulk}}  \left(D_{\textrm{gas}} n_{\textrm{gas}}\right) \left\langle Y_\textrm{net} v \right\rangle_\textrm{skM}, \label{tsput}
\end{align}
(\citealt{Barlow1978,Draine1979, Tielens1994, Nozawa2006, Bocchio2014}) is the reduction of grain radius per unit time, $\mu_\textrm{dust}=20$ is the mean atomic weight of silicate grain atoms, $m_\textrm{amu}$ is the atomic mass unit, $\rho_\textrm{bulk}$ is the dust bulk density, $n_\textrm{gas}$ is the gas number density, and $\left\langle Y_\textrm{net} v \right\rangle_\textrm{skM}$ is the product of net yield and velocity $v$ averaged over the skewed Maxwellian velocity distribution. We introduce here a gas depletion factor 
\begin{align}
 D_{\textrm{gas}} &=  \frac{\tau_\textrm{d}}{\Delta t}\left(1-\exp{\left[-\frac{\Delta t}{\tau_\textrm{d}}\right)}\right], \hspace*{0.5cm} D_{\textrm{gas}}\in\left[0,1\right]\label{depl}\\
 \textrm{with}\nonumber\\
 \tau_\textrm{d}&=\left(\pi a^2 n_a \left\langle -\left(Y_\textrm{acc}+Y_\textrm{trap}\right) v \right\rangle_\textrm{skM}\right)^{-1} \label{eq_dep}
\end{align}
as the gas depletion timescale and $n_a$ as the number density of dust grains with radius $a$. Considering a typical number density $n_a=\unit[0.01]{m^{-3}}$ for grains with radius $a=\unit[0.1]{\mu m}$ (for a clump gas density of $n_\textrm{cl}=\unit[100]{cm^{-3}}$, a gas-to-dust mass ratio \mbox{$\Delta_\textrm{gd}=10$} and an initial log-normal distribution with \mbox{$a_\text{peak}=\unit[0.1]{\mu m}$} and \mbox{$\sigma=0.1$}), \mbox{$\left(Y_\textrm{acc}+Y_\textrm{trap}\right)\approx-1$,}
and a velocity of the dust relative to the gas of the order of $v=\unit[10]{km/s}$,
the gas depletion timescale amounts to $\tau_\textrm{d}\unit[ \approx 10^4]{yr}$. However, as small dust grains are generated due to dust fragmentation, the number density of $\unit[1]{nm}$ grains in shocked and compressed clumps can take values up to the order of  $n_a\unit[\sim10^6]{m^{-3}}$, and the resulting depletion timescale is then only $\tau_\textrm{d}\unit[\approx 1]{yr}$ which is comparable to the
used time-step $\Delta t = \unit[0.5]{yr}$. The high post-shock gas temperatures cause even higher velocities $v$ and thus reduce the gas depletion time-scale which becomes crucial for the further evolution.

For regular sputtering, the gas particles are not trapped or accreted \mbox{($Y_\textrm{acc}=Y_\textrm{trap}=0$)} and the gas number density $n_\textrm{gas}$ is unaffected so that the depletion factor is 1. However, when gas accretion or trapping are present, oxygen atoms are removed from the gas and the gas number density decreases during the time-step $\Delta t$. To take that into account,  the actual gas number density in equation~(\ref{tsput}) is $D_{\textrm{gas}} n_{\textrm{gas}}$ (see Appendix \ref{app_depl} for a derivation of $D_\textrm{gas}$). 

Due to the nature of the post-processing we are not able to reduce the gas number density from one time-step to the next as this would strongly affect the hydrodynamical simulations. Therefore, we are limited to considering gas depletion only for the calculation of the yields and we have to neglect the evolution of gas depletion on a longer timescale. Moreover, we also neglect the destroyed dust material which contributes to the regular gas and which was introduced by K19 as ``dusty gas''. Test simulations have shown that the production of new grain material by ion trapping together with destruction of the material by sputtering or grain-grain collisions could lead under extreme conditions to an unrealistically high dusty gas density which would at least at the end of our simulations completely dominate the evolution of both the gas and dust component. The main reason for this high dusty gas density is the fact that we cannot follow the depletion of the regular gas. Therefore, the gas would serve as an inexhaustible source of material resulting in unrealistically high gas trapping and accretion rates. On the other hand, we expect that the disregard of the dusty gas component counters the continuous refreshing of the regular gas component and that both effects cancel each other out: when dust destruction by sputtering and dust growth by ion trapping and accretion are at similar levels, the sum of the gas mass and the dust grain mass is fixed.


\section{Results}
\label{sec3}

\begin{figure}
 \centering
  \includegraphics[trim=2.2cm 3.71cm 0.9cm 2.1cm, clip=true,width=1.0\linewidth, page=1]{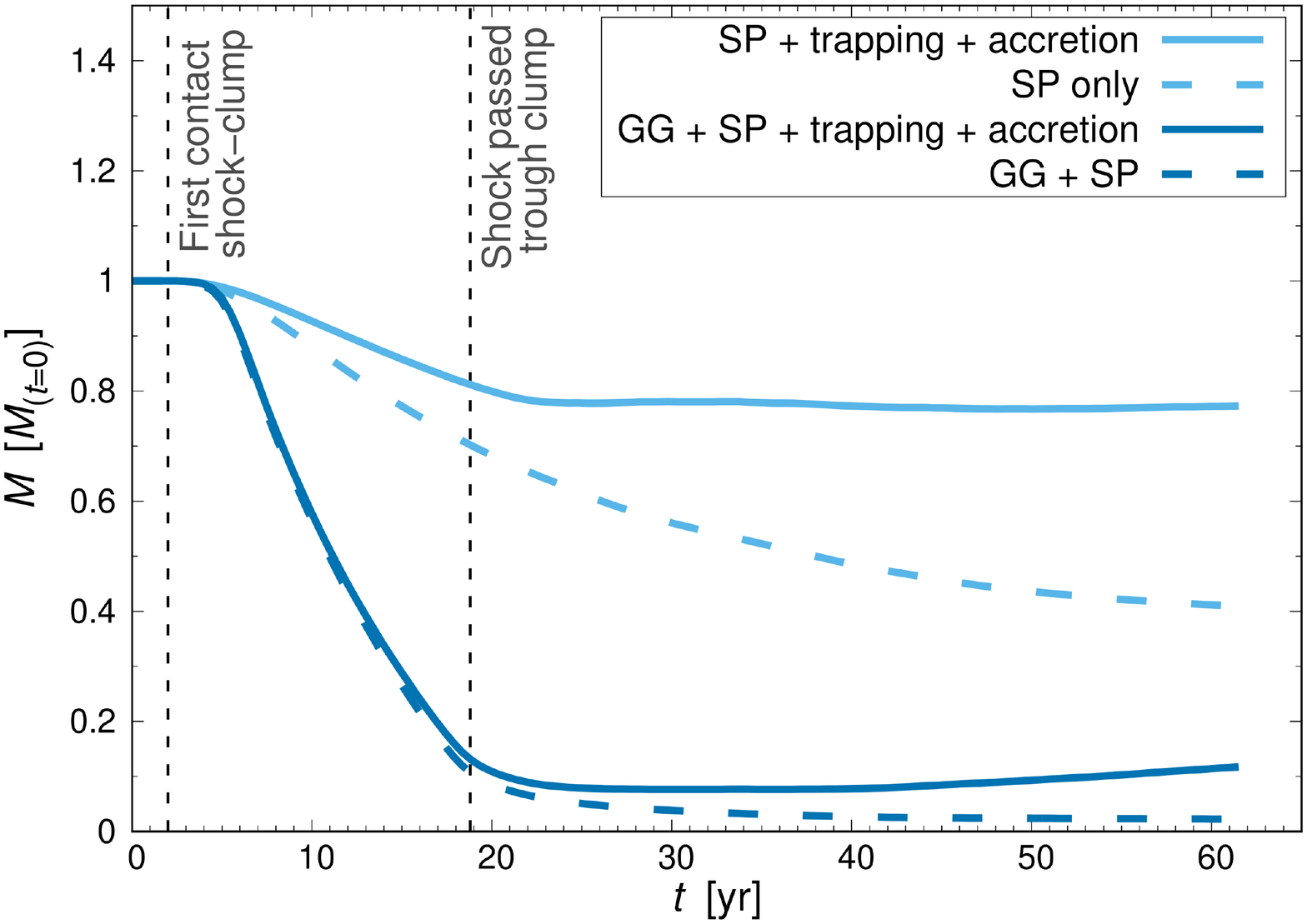}\\[-0.715cm]
  \includegraphics[trim=2.525cm 0cm 0.9cm 0.3cm, clip=true,width=1.0\linewidth, page=2]{Pics/Evolution_total.pdf} 
   \caption{\textit{Top}: Surviving dust mass $M$ of the clump in units of the initial clump dust mass $M_\textrm{(t=0)}$ as a function of time, taking into account grain-grain collisions (GG) and sputtering (SP) with or without trapping or accretion of oxygen ions (different colors and line types). Oxygen trapping and gas accretion reduce the dust destruction significantly. \textit{Bottom}: Dust destruction and growth rates as a function of time. Both sputtering and grain-grain collisions are considered for the dust destruction  while the dust growth is given by the mass of trapped or accreted oxygen per unit time. The difference between these two rates determines the total dust mass evolution in the \textit{top} panel (solid dark-blue line). The time points when the shock has first contact with the clump as well as when it has travelled through the clump are indicated in both panels as vertical dashed lines.}
   \label{res_dust_evo100} 
\end{figure} 

      \begin{figure}
 \centering
  \includegraphics[trim=2.5cm 2.1cm 0.9cm 2.1cm, clip=true,width=1.0\linewidth, page=1]{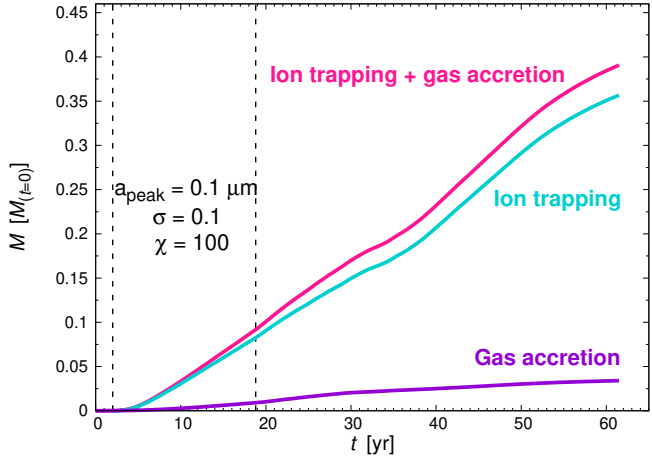} 
   \caption{Dust mass $M$ (in units of the initial clump dust mass $M_\textrm{(t=0)}$) gained by ion trapping and/or accretion of the gas as a function of time. The plots do not take into account that a significant part of the newly produced dust masses could be destroyed again by sputtering or grain-grain collisions.}
   \label{fig_growth} 
\end{figure}

    \begin{figure*}
  \centering
   \includegraphics[trim=0cm 0.0cm 0.0cm 0.0cm, clip=true,width=0.915\linewidth, page=1]{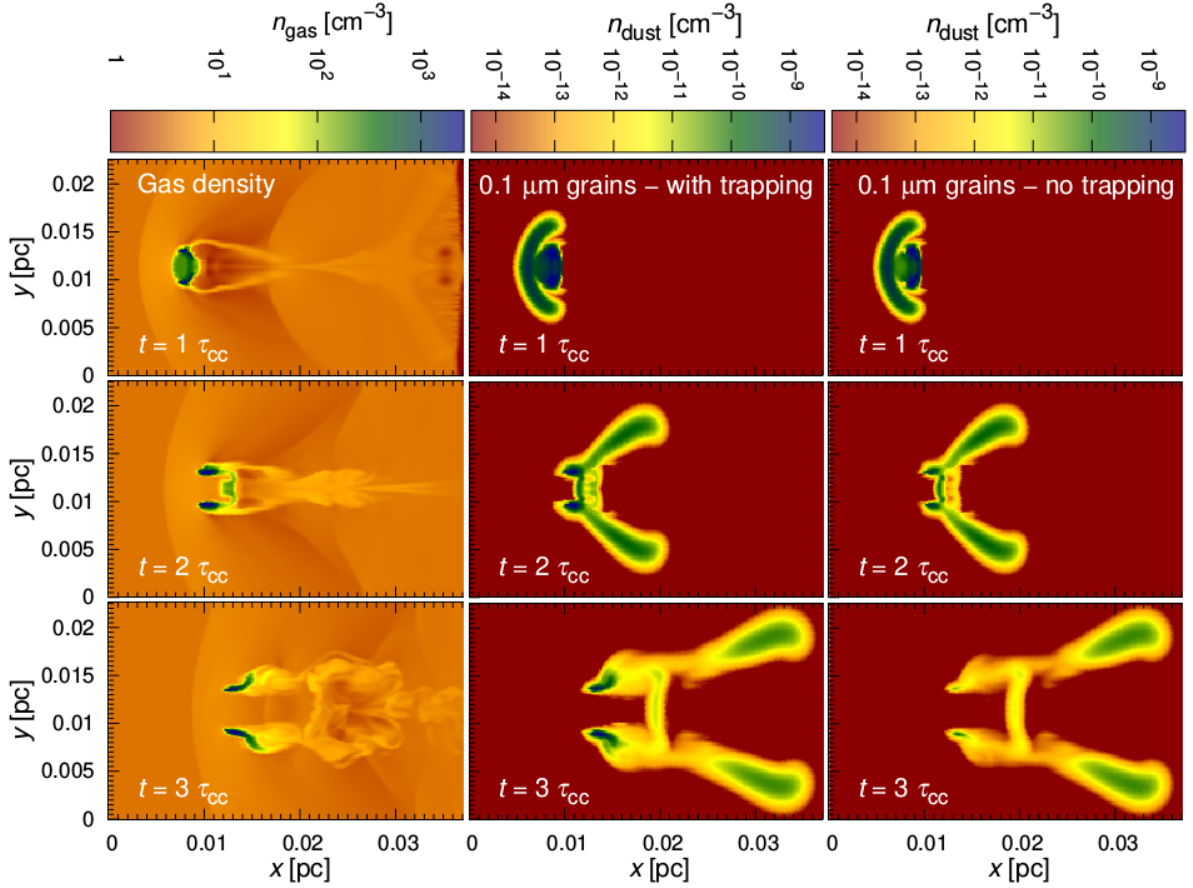}
  \caption{Temporal evolution of the spatial gas density (\textit{left} column) as well as the number density of $\unit[0.1]{\mu m}$ dust grains, with (\textit{centre}) and without oxygen trapping (\textit{right}). The rows show the distributions at time $t=1, 2$, and $3$ cloud-crushing times $\tau_\textrm{cc}$, respectively, after the first contact of the reverse shock with the clump. The panels show a fixed cutout of the computational domain and the colour scale is fixed for each figure column.}
 \label{fig_maps} 
   \end{figure*}

We conducted 3D hydrodynamical simulations of the cloud-crushing problem using \textsc{AstroBEAR} and determined the integrated dust mass  as a function of time using \mbox{\textsc{Paperboats}}. 

The impact of ion trapping and gas accretion on the total dust survival rate can be clearly seen in Fig.~\ref{res_dust_evo100} (\textit{top}). 
When the shock impacts the clump,  the  combined effects of grain-grain collisions and sputtering rapidly decrease the total dust mass. The destruction weakens after ${\sim}1$ cloud-crushing time (${\sim}\unit[20]{yrs}$) and the remaining dust mass is mostly maintained. Taking ion trapping and gas accretion into account, the total dust mass even starts to increase slightly after the first cloud-crushing time and reaches ${\sim}\unit[12]{\%}$ of the original mass after three cloud-crushing times. For comparison, when ion trapping and accretion are neglected the dust survival rate decreases during the entire simulation time and ends up at only ${\sim}\unit[3]{\%}$. Neglecting grain-grain collisions, ${\sim}\unit[80]{\%}$ of the initial dust mass survives for the case of sputtering combined with ion trapping while the survival rate is ${\sim}\unit[40]{\%}$ for sputtering without ion trapping and accretion. Ion trapping and gas accretion increase the surviving dust mass fraction.
 
Fig.~\ref{res_dust_evo100} (\textit{bottom}) shows the dust growth rate $\dot{M}_\textrm{growth}\left(t\right)$ due to oxygen ion trapping and accretion as well as the rate of destruction $\dot{M}_\textrm{destr}\left(t\right)$ through sputtering and grain grain collisions. The difference between them determines the total dust mass via $M(t+\Delta t)=M(t)+\left(\dot{M}_\textrm{growth}\left(t\right)-\dot{M}_\textrm{destr}\left(t\right)\right)\Delta t$. In the first cloud-crushing time, the shock travels through the clump causing high temperatures and velocities which result in a high destruction rate. Afterwards, the clump decays and small-scale structures (of the order of ${\sim}\unit[1]{mpc}$) with high gas densities are formed. The dust material is concentrated in these gas clumps, forming high dust density structures: dust growth as well as destruction processes are roughly at the same level in this environment. Dust growth slightly surpasses dust destruction at later times, which is one of the few differences between the 3D and 2D simulations: for the latter, the dust destruction and growth rates flatten after $\unit[{\sim}50]{yr}$ (not shown).

Oxygen  trapping is the dominant growth process compared to gas accretion (Fig.~\ref{fig_growth}). The mass gained by ion trapping increases linearly with time and reaches $\unit[{\sim}35]{\%}$ of the initial clump dust mass $M_\textrm{(t=0)}$ after three cloud-crushing times. In comparison, the newly produced dust mass due to gas accretion amounts to $0.03\,M_\textrm{(t=0)}$ at the end of the simulation. At around two cloud-crushing times, ion trapping is reinforced due to the reduced relative velocities between gas and dust components, and the mass gained by ion trapping rises slightly stronger. In total, ion trapping makes up $\unit[{\sim}90]{\%}$ of the newly produced dust mass after three cloud-crushing times. The accreted gas material will be predominantly located close to the surface of the grains where it can be more easily destroyed by, e.g., sputtering than trapped atoms inside the grains. Thus the contribution of the ion trapping to the dust mass gain is even higher.

Fig.~\ref{fig_maps} shows the spatial distribution of the gas and of $\unit[0.1]{\mu m}$ silicate grains with or without oxygen trapping and accretion for $1, 2,$ and $3$ cloud-crushing times after the shock has impacted the clump. The maps represent the central cut of the 3D domain through the initial center of the clump. In the case of ion trapping and accretion the dust component has higher densities but also higher density contrasts (over-dense knots) compared to the case of no ion trapping or accretion. The equivalent maps in 2D (not shown) have gas and dust density distributions which are  less evolved in the shock direction but broader perpendicular to the shock direction. We emphasize that the maps in Fig.~\ref{fig_maps} only depict the distribution of $\unit[0.1]{\mu m}$ radius grains but that due to the dust processing, grains of much smaller or larger radius (at least for the case of oxygen trapping) can be formed. Consequently, the maps for other grain sizes are significantly different and indicate regions of enhanced rates of dust production or destruction.

    \begin{figure}
 \centering
  \includegraphics[trim=2.4cm 2.1cm 2.1cm 2.5cm, clip=true,width=1.0\linewidth, page=1]{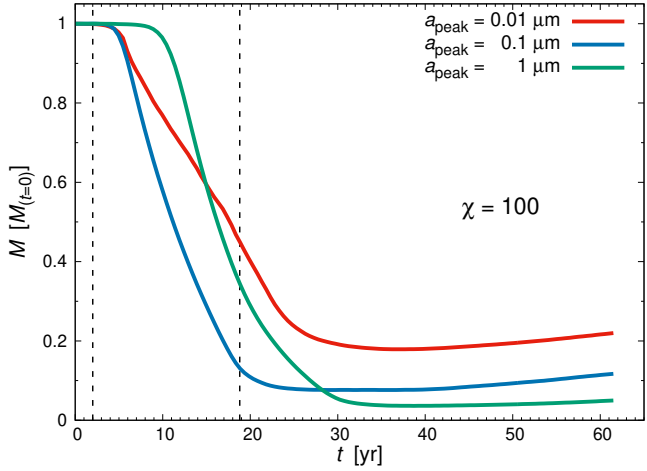} 
   \caption{Surviving dust mass $M$ in a clump in units of the initial clump dust mass $M_\textrm{(t=0)}$ as a function of time for different initial log-normal grain size distributions, given by peak radii $a_\textrm{peak}=\unit[0.01]{\mu m}$ (red solid line), $\unit[0.1]{\mu m}$ (blue), and $\unit[1]{\mu m}$ (green), and width of the initial distribution $\sigma=0.1$. Grain-grain collisions and sputtering with oxygen ion trapping and gas accretion are considered.}
   \label{res_dust_evo3} 
\end{figure}

 A comparison between simulations for different initial log-normal grain size distributions with $a_\textrm{peak}=0.01, 0.1$ and $\unit[1]{\mu m}$ (Fig.~\ref{res_dust_evo3}) shows that the surviving dust mass depends strongly on the initial size distribution of the grains ($\unit[22]{\%}$, $\unit[12]{\%}$, $\unit[5]{\%}$, resp.). This is because the effect of oxygen ion trapping and accretion is a function of grain size which can be explained by the dependence of the depletion timescale $\tau_\textrm{d}$ on the grain radius for a fixed dust-to-gas mass ratio. The effect of the grain size on the quantity $\left\langle -\left(Y_\textrm{acc}+Y_\textrm{trap}\right) v \right\rangle_\textrm{skM}$ is negligible and the timescale  for gas depletion is then proportional to $(a^2 n_a)^{-1}$ (equation~\ref{eq_dep}). As the initial dust mass is fixed, it follows $n_a\propto a^{-3}$ and $\tau_\textrm{d}\propto a$. The larger the initial grain sizes, the smaller are the effects of ion trapping and gas accretion and thus the surviving dust mass.\footnote{We note, that this is a consequence of the fixed initial dust mass in the clump, which results in a larger dust cross section (integrated over all particles) the smaller the dust grain size. Furthermore, the initial dust grains must have a minimum grain radius in order to survive the impact of the shock during the first cloud-crushing time.
 } 
 
   \begin{figure*} 
 \centering
  \includegraphics[trim=0.0cm 0.0cm 0.0cm 0.0cm, clip=true,width=1.0\linewidth, page=1]{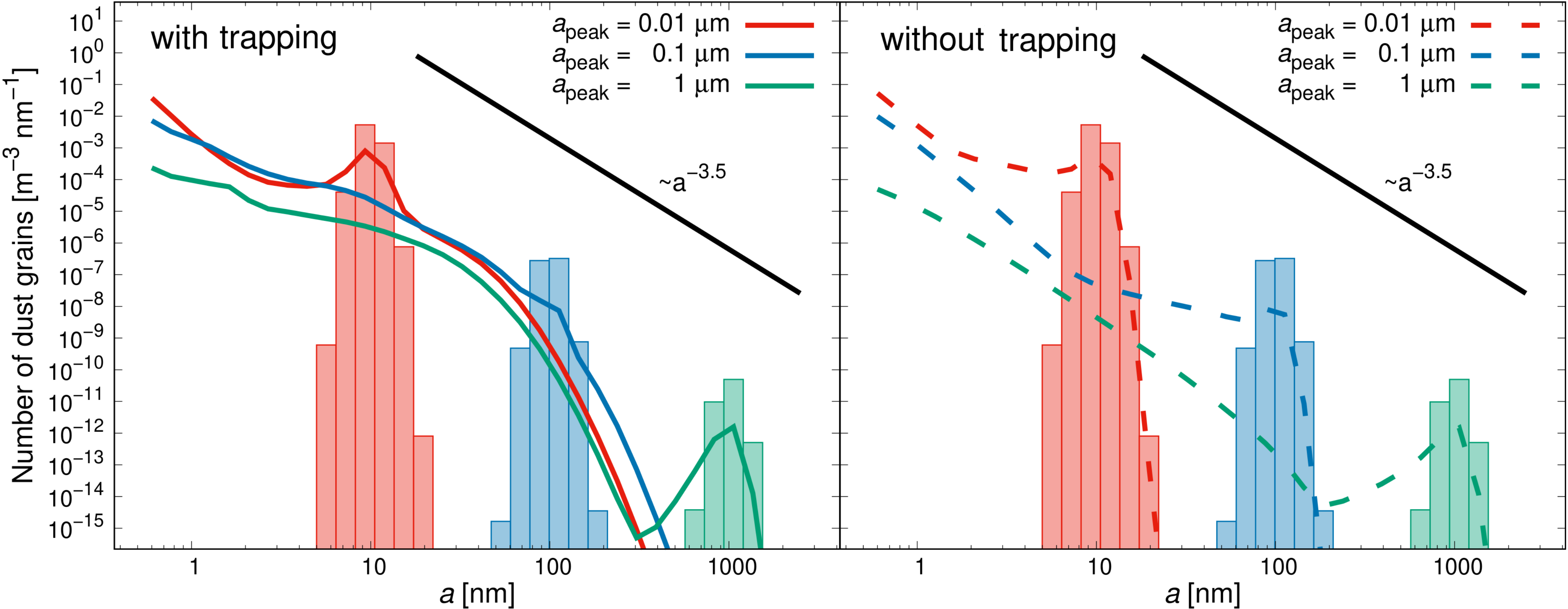}  
   \caption{Final grain size distributions for initial log-normal distributions with $a_\textrm{peak}=0.01, 0.1$ and $\unit[1.0]{\mu m}$ (red, blue, and green lines, respectively; density contrast $\chi=100$) and distribution width $\sigma=0.1$, with (\textit{left}) and without (\textit{right}) ion trapping or accretion. The initial distributions are shown as histograms and the slope of the common $\gamma=-3.5$ power-law is also shown.}
   \label{res_fin_distrib} 
\end{figure*} 
 
\begin{figure}
 \centering
  \includegraphics[trim=2.2cm 2.1cm 0.9cm 2.1cm, clip=true,width=0.99\linewidth, page=1]{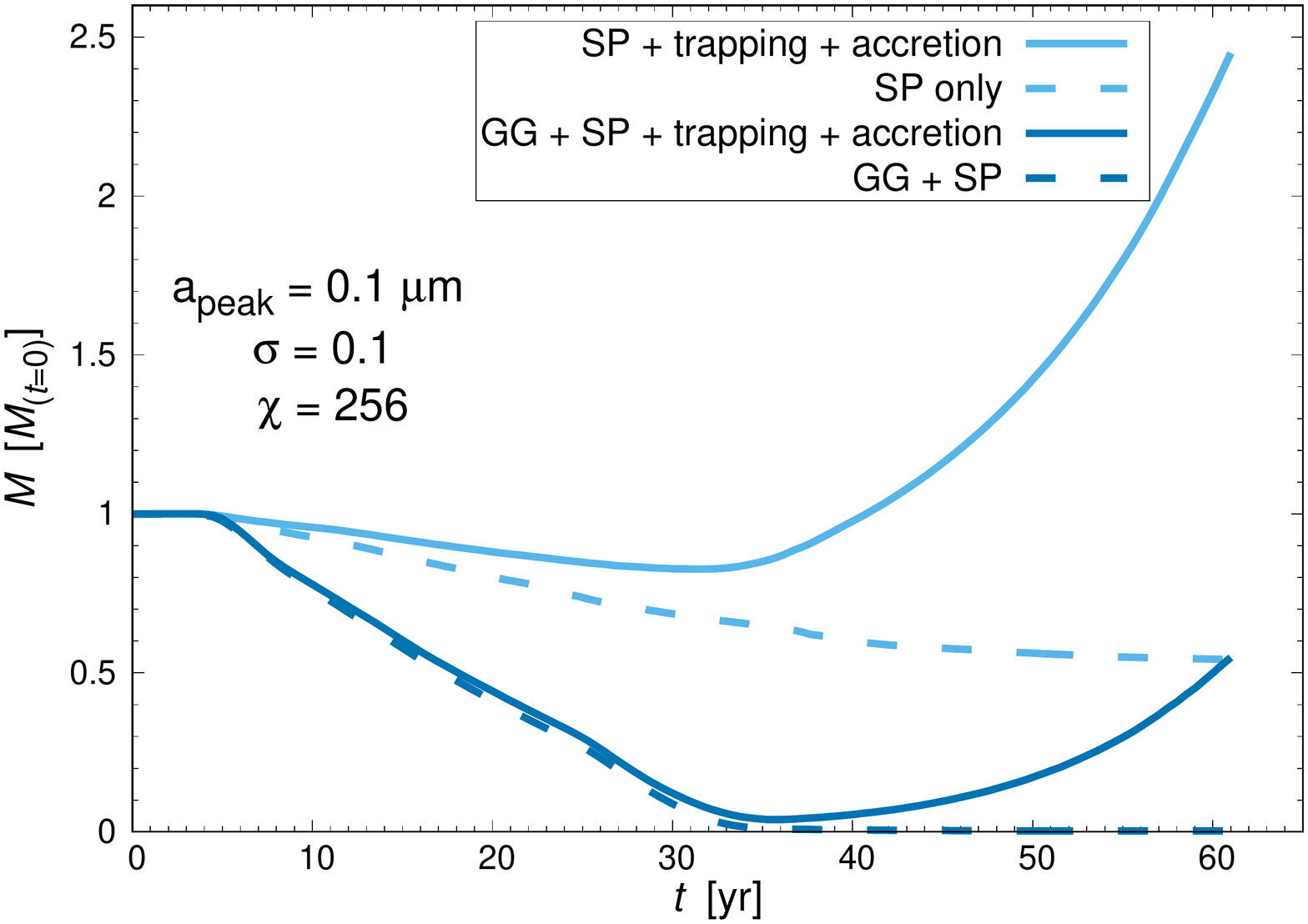}
  \caption{Same as Fig.~\ref{res_dust_evo100}  (\textit{top}), but for $\chi=256$.}
   \label{res_dust_evo100_256} 
 \centering
  \includegraphics[trim=2.5cm 2.1cm 0.9cm 2.1cm, clip=true,width=0.99\linewidth, page=1]{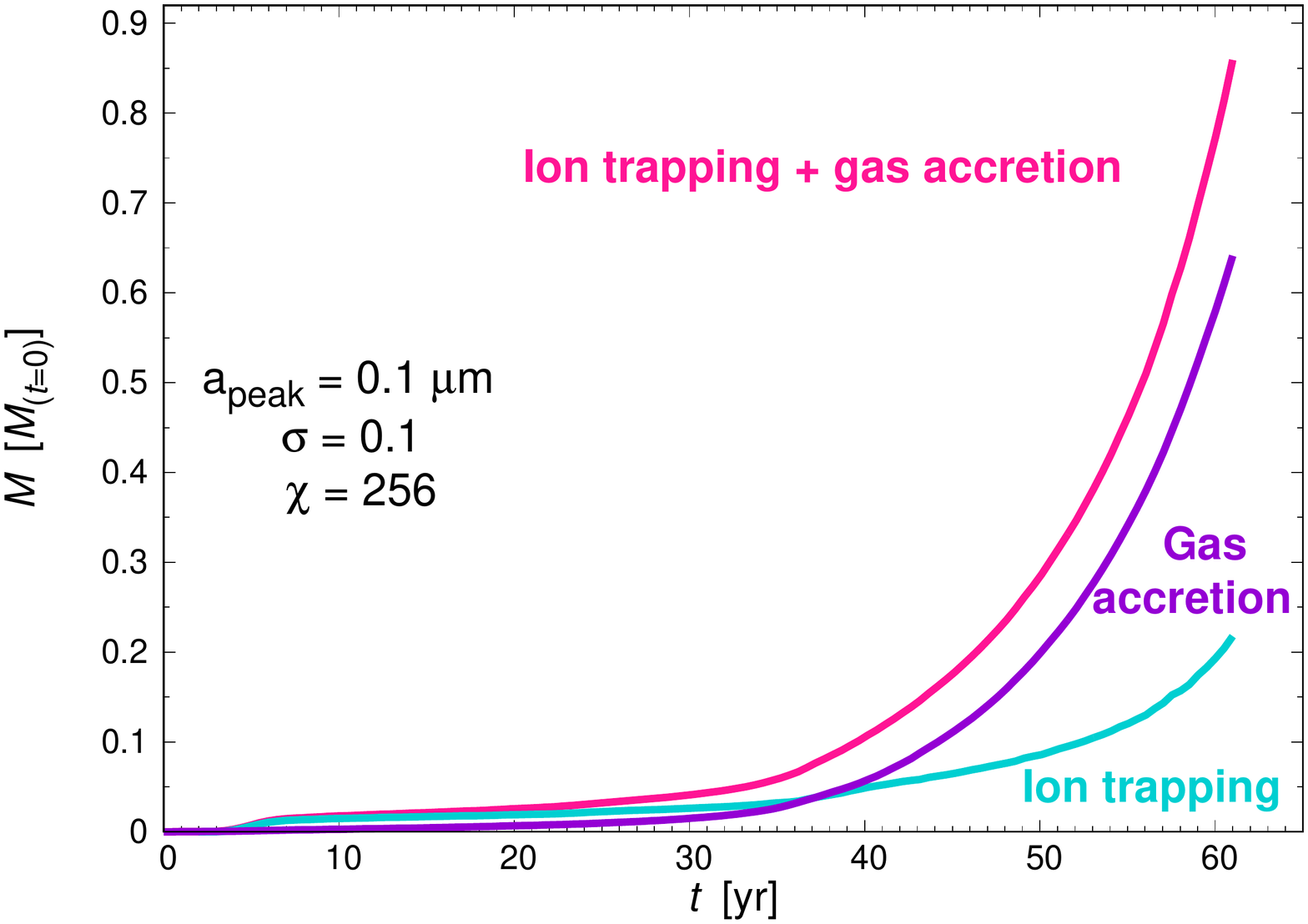} 
   \caption{Same as Fig.~\ref{fig_growth}, but for $\chi=256$. The plots do not take into account that a significant part of the newly produced dust masses could be destroyed again by sputtering or grain-grain collisions.}
   \label{fig_growth_256} 
   \end{figure} 
 
The dust mass of all three distributions is primarily destroyed within the first cloud-crushing time before ion trapping and accretion mitigate the destruction and even starts to increase the dust mass. Two effects are visible in Fig.~\ref{res_dust_evo3}: oxygen ion trapping and accretion are more efficient for smaller grain sizes, as well as the fact that the evolution of larger grains is delayed compared to small grains. Grain destruction in the simulations for $a_\textrm{peak}=\unit[1]{\mu m}$ grains starts $\unit[{\sim}5]{yr}$ after the destruction begins for initial distributions with $a_\textrm{peak}=0.01$ and $\unit[0.1]{\mu m}$, as grain-grain collisions are the main destruction mechanism for these large grains and their stopping time, and thus their time to be accelerated, is of the order of years.
 
The dust processing completely re-distributes the grain size distribution after three cloud-crushing times (Fig.~\ref{res_fin_distrib}, \textit{left}). On the one hand, destruction processes produce smaller grains compared to the initial grain size distribution, on the other hand, dust growth processes result in an increase of larger dust grains. K19 found that the final grain size distribution is composed of two components --  a remnant of the initial distribution, located around $a_\textrm{peak}$ though reduced in grain number density due to sputtering and collisions, plus a power-law distribution of smaller dust grains that reflect the fragments of shattering collisions. Through ion trapping and gas accretion the power-law distribution is extended to larger grains than $a_\textrm{peak}$, at least for the simulations with $a_\textrm{peak}=0.01$ and $\unit[0.1]{\mu m}$ for which oxygen trapping and accretion are efficient. We show as a comparison in Fig.~\ref{res_fin_distrib} (\textit{right}) the final grain size distribution for simulations without oxygen trapping or gas accretion, clearly indicating a reduced number of large grains (surviving rates of $\unit[12]{\%}$, $\unit[3]{\%}$ and $\unit[3]{\%}$ for $a_\textrm{peak}=0.01, 0.1$ and $\unit[1]{\mu m}$, resp.).

For a density ratio of $\chi=100$, the velocity of the reverse shock within the clump amounts to $\unit[{\sim}160]{km/s}$. The kinetic energy of a  $\unit[160]{km/s}$ oxygen ion is of the order of $\unit[2.1]{keV}$ which corresponds to a net destruction of grain material (Fig.~\ref{res_dust_evo}). In contrast, a density ratio\footnote{An oxygen ion with velocity  $\unit[100]{km/s}$ has a kinetic energy of $\unit[0.84]{keV}$, at which the net yield is negative in the case of ion trapping (Fig.~\ref{res_dust_evo}, \textit{right}). The grain effectively grows for each ion that impacts with that or lower velocity. The shock in the ambient medium has a velocity of $v_\textrm{sh}=\unit[1600]{km/s}$ and is decelerated in the clump to $v_\textrm{sh,cl}=\nicefrac{v_\textrm{sh}}{\sqrt{\chi}}$. The oxygen ions can thus reach velocities up to $\unit[100]{km/s}$ when the density contrast amounts to \mbox{$\chi=\left(\nicefrac{v_\textrm{sh}}{v_\textrm{sh,cl}}\right)^2= \left(\nicefrac{1600}{100}\right)^2=256$.}}
of $\chi=256$ causes a decrease of the shock velocity to $\unit[100]{km/s}$ within the clump, and oxygen ions of that velocity have a kinetic energy of $\unit[0.84]{keV}$ which corresponds to a net gain of grain material. Consequently, we expect a more effective ion trapping mechanism as well as a larger surviving dust mass for the higher density contrast.

\begin{figure}
  \includegraphics[trim=2.4cm 2.1cm 2.1cm 2.5cm, clip=true,width=1.0\linewidth, page=1]{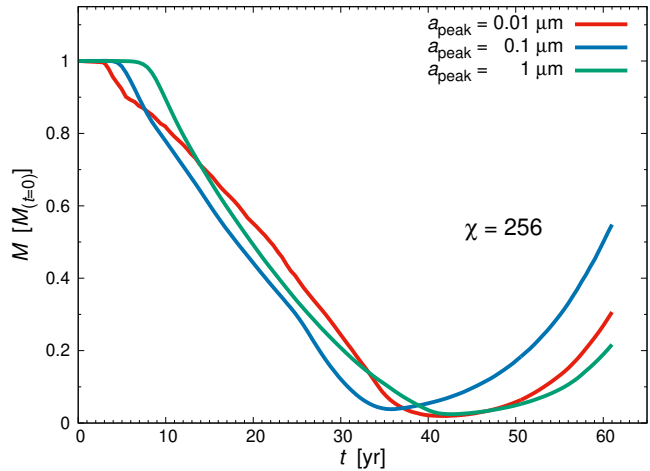} 
   \caption{Same as Fig.~\ref{res_dust_evo3}, but for $\chi=256$.}
   \label{res_dust_evo3_256} 
\end{figure}

We conducted hydrodynamical simulations followed by dust post-processing for the density contrast $\chi=256$ \mbox{(Figs.~\ref{res_dust_evo100_256}-\ref{res_dust_evo3_256}).} As expected, the dust survival rates are larger than for \mbox{$\chi=100$}. For $a_\text{peak}=\unit[0.1]{\mu m}$, $\unit[{\sim}55]{\%}$ of the initial dust mass survives when ion trapping and gas accretion are taken into account ($\unit[{\sim}12]{\%}$ for $\chi=100$), while the survival rate is below $\unit[1]{\%}$ without these mechanisms ($\unit[{\sim}3]{\%}$ for $\chi=100$). 

Grain-grain collisions destroy only a minor part of the dust material (K19) but cause the production of smaller fragments. As the survival rate drops below $\unit[10]{\%}$ in the first $\unit[{\sim}35]{yrs}$, the grains must be destroyed by sputtering. In the case of ion trapping, the net yield at $\unit[0.84]{keV}$ is negative (see~Fig.~\ref{res_dust_evo}) which implies dust growth for each impinging oxygen ion. However, not all ions can be trapped as some instead tunnel through the small dust grains created by grain-grain collisions.

Fig.~\ref{fig_growth_256} shows that the total mass gained is about $\unit[{\sim}90]{\%}$ of the initial clump dust mass ($\chi=256$), which is twice the gained mass of the density contrast $\chi=100$. However, gas accretion makes a larger contribution ($\unit[74]{\%}$) than ion trapping ($\unit[26]{\%}$). The reason is that  $\unit[{\sim}100]{km/s}$ is the maximum velocity of the oxygen ions, which predominantly occurs at the moment when the reverse shock impacts the clump and hits the unaccelerated dust grains. At later time points, the average velocities are lower and the energies of the oxygen ion are not sufficient for ion trapping, but  instead favour gas accretion. Therefore, the efficiency of ion trapping is also strongly dependent on the clump density and the shock velocity. Furthermore, the different gas density has an effect on the grain-size dependence of the dust growth processes. Fig.~\ref{res_dust_evo3_256} shows that initial grain distributions with $a_\text{peak}=\unit[0.1]{\mu m}$ have the highest survival rates for the density contrast $\chi=256$.

\begin{figure}
  \includegraphics[trim=2.4cm 2.1cm 2.1cm 2.5cm, clip=true,width=1.0\linewidth, page=1]{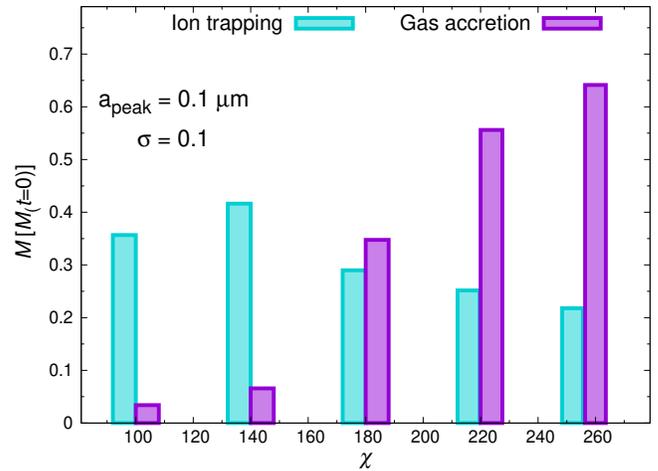} 
   \caption{Gained dust mass due to oxygen trapping and gas accretion as a function of density contrast $\chi$.}
   \label{res_dust_evo_all} 
\end{figure}

Gas accretion makes a larger contribution to the gained mass than ion trapping when the density contrast is $\chi=256$, while the situation is the opposite for $\chi=100$. Consequently, there exists a critical density contrast at which gas accretion becomes more important than ion trapping. We conducted hydrodynamical simulations followed by dust post-processing (initial log-normal grain size distribution with \mbox{$a_\textrm{peak}=\unit[0.1]{\mu m}$}, \mbox{$\sigma=0.1$}) for the density contrasts $\chi=140, 180$ and $220$ and determined the dust mass gained by ion trapping and gas accretion \mbox{(Fig.~\ref{res_dust_evo_all}).} With increasing density contrast the dust mass gained due to oxygen trapping is in a rough trend decreasing while the accreted mass increases. At the density contrast $\chi\approx180$ the mass gains from both effects are at a similar level (${\sim}0.3\,M_\textrm{(t=0)}$). As gas acrretion and ion trapping are both functions of the grain size, the critical density contrast depends on the initial grain size distribution.
 
\section{Conclusions}
\label{sec4}
We have investigated the influence of trapping of oxygen and other ions on the survival rate of dust grains through the passage of the reverse shock in the ejecta of the SNR Cas~A. We found that oxygen trapping has the potential to significantly increase the surviving dust mass. In the case of a clump to ambient medium gas density contrast of $\chi=100$ and an initial log-normal grain size distribution with peak radius $a_\textrm{peak}=\unit[0.01]{\mu m}$ ($\unit[0.1]{\mu m}, \unit[1]{\mu m}$), the silicate survival rate is $\unit[12]{\%}$ ($\unit[3]{\%}, \unit[3]{\%}$) when oxygen trapping and gas accretion are neglected, but is $\unit[22]{\%}$ ($\unit[12]{\%}, \unit[5]{\%}$) when these effects are taken into account. For larger density contrasts ($\chi\gtrsim180$ for $a_\textrm{peak}=\unit[0.1]{\mu m}$) the efficiency of gas accretion can surpass that of ion trapping in generating new dust material, resulting in higher survival rates.

The  dust grain growth processes can produce a significant amount of grains that have sizes above the initial grain sizes. Therefore, ion trapping and gas accretion can play a crucial role in the surviving dust budget of oxygen-rich SNRs and thereby contribute to the dust budget in the interstellar medium.  Finally, the increased grain sizes might account for the discrepancy between observed and theoretically predicted dust grain sizes in some SNRs.


\section*{Acknowledgements}
FK, MJB and FS acknowledge funding from the European Research Council Grant SNDUST ERC-2015-AdG-694520. We would like to thank Erica Fogerty for providing generous support in the application of AstroBEAR.
Simulations were performed using the UCL HPC RC cluster {\textsc{GRACE}} and the data intensive {\textsc{Peta4-Skylake}} service at Cambridge. {\textsc{Peta4-Skylake}} usage is supported through DiRAC project ACSP190 (SNDUST) using the Cambridge Service for Data Driven Discovery (CSD3), part of which is operated by the University of Cambridge Research Computing on behalf of the STFC DiRAC HPC Facility (\href{www.dirac.ac.uk}{www.dirac.ac.uk}). The DiRAC component of CSD3 was funded by BEIS capital funding via STFC capital grants ST/P002307/1 and ST/R002452/1 and STFC operations grant ST/R00689X/1. DiRAC is part of the National \mbox{e-Infrastructure.} We thank Clare Jenner (UCL), Lydia Heck (Durham University), UCL RC support, and Cambridge HPCS support for their assistance.

\software{Paperboats \citep{Kirchschlager2019b},  
          AstroBEAR \citep{Carroll-Nellenback2013},
          UCL HPC RC cluster GRACE,
          Peta4-Skylake (DiRAC project;\\ \href{www.dirac.ac.uk}{www.dirac.ac.uk})
          }

  \bibliographystyle{aasjournal}
{\footnotesize
  \bibliography{Oxy_trapping}
}

\appendix

\section{Derivation of the gas depletion factor} 
\label{app_depl}
We present here the derivation for the gas depletion factor $D_\textrm{gas}$ (equation~\ref{depl}).
 Due to ion trapping and gas accretion, the gas number density $\tilde{n}_\textrm{gas}(t)$ decreases during each time-step $\Delta t$ as an e-folding function of time,
$\tilde{n}_\textrm{gas}(t)= \exp{\left[-\frac{t}{\tau_\textrm{d}}\right]} n_\textrm{gas}$, 
with $n_\textrm{gas} = \tilde{n}_\textrm{gas}(0)$ as initial gas density at each time-step and $\tau_\textrm{d}$ as the gas depletion timescale (equation~\ref{eq_dep}). To take into account the changing gas density during $\Delta t$ for the sputtering and trapping processes, we introduce the gas depletion factor $D_\textrm{gas}$ which is the average gas density during the time interval $[0,\Delta t]$ divided by $n_\textrm{gas}$. Following the mean value theorem, $D_\textrm{gas}n_\textrm{gas}$ is  the integral of $\tilde{n}_\textrm{gas}(t)$ with bounds 0 and $\Delta t$ divided by the length of the interval,
  \begin{align}
  D_\textrm{gas} &=\frac{1}{\Delta t}\int\limits_0^{\Delta t} \exp{\left[-\frac{t'}{\tau_\textrm{d}}\right]} \textrm{d}t' =  \left(1 - \exp{\left[-\frac{\Delta t}{\tau_\textrm{d}}\right]} \right) \frac{\tau_\textrm{d}}{\Delta t}.\nonumber
   \end{align}
   
We note that the gas depletion factor $D_\textrm{gas}$ can also be derived in a second approach taking into account 
mass conservation between the reduced gas mass on the one hand and the increase of dust mass that is generated due to gas accretion and trapping on the other hand.

\end{document}